\documentclass[twocolumn,showpacs,showkeys,nofootinbib,pre]{revtex4}
\usepackage{graphicx}
\usepackage{soul}
\usepackage{color}
\usepackage{amsmath}
\usepackage{stackrel}
\begin{document}
\title{Multistable Dissipative Breathers and Novel Collective States in SQUID 
 Lieb Metamaterials}  
\author{N. Lazarides$^{1,2,3}$, G. P. Tsironis$^{1,2,3,4}$}
\affiliation{
$^{1}$Department of Physics, University of Crete, P. O. Box 2208, 
      71003 Heraklion, Greece; \\
$^{2}$Institute of Electronic Structure and Laser, Foundation for Research and 
      Technology--Hellas, P.O. Box 1527, 71110 Heraklion, Greece \\
$^{3}$National University of Science and Technology "MISiS", Leninsky prosp. 4, 
      Moscow, 119049, Russia; \\ 
$^{4}$School of Engineering and Applied Sciences,  Harvard University, 
      Cambridge, Massachusetts 02138, USA
}
\date{\today}
\begin{abstract}
A SQUID (Superconducting QUantum Interference Device) metamaterial on a Lieb 
lattice with nearest-neighbor coupling supports simultaneously stable 
dissipative breather families which are generated through a delicate balance of 
input power and intrinsic losses. Breather multistability is possible due to the 
peculiar snaking flux ampitude - frequency curve of single dissipative-driven 
SQUIDs, which for relatively high sinusoidal flux field amplitudes exhibits 
several stable and unstable solutions in a narrow frequency band around 
resonance. These breathers are very weakly interacting with each other, while 
multistability regimes with different number of simultaneously stable breathers 
persist for substantial intervals of frequency, flux field amplitude, and 
coupling coefficients. Moreover, the emergence of chimera states as well as 
novel temporally chaotic states exhibiting spatial homogeneity within each 
sublattice of the Lieb lattice is demonstrated.
\end{abstract}
\maketitle
\section{Introduction}
The {\em superconducting metamaterials}, a particular class of artificial 
mediums which relay on the sensitivity of the superconducting state reached by 
their constituting elements at low temperatures, have recently been the focus of
considerable research efforts \cite{Anlage2011,Jung2014,Lazarides2018}. The 
superconducting analogue of conventional (metallic) metamaterials, which can 
become nonlinear with the insertions of appropriate electronic components 
\cite{Lapine2003,Lapine2014}, are the SQUID (Superconducting QUantum 
Interference Device) metamaterials. The latter are inherently nonlinear due to 
the Josephson effect \cite{Josephson1962}, since each SQUID, in its simplest 
version, consists of a superconducting ring interrupted by a Josephson junction. 
The concept of SQUID metamaterials was theoretically introduced more than a 
decade ago both in the quantum \cite{Du2006} and the classical 
\cite{Lazarides2007} regimes. Recent experiments on SQUID metamaterials have 
revealed several extraordinary properties such as negative diamagnetic 
permeability \cite{Jung2013,Butz2013a}, broad-band tunability 
\cite{Butz2013a,Trepanier2013}, self-induced broad-band transparency 
\cite{Zhang2015}, dynamic multistability and switching \cite{Jung2014b}, as well 
as coherent oscillations \cite{Trepanier2017}. Moreover, nonlinear localization 
\cite{Lazarides2008a} and nonlinear band-opening (nonlinear transmission) 
\cite{Tsironis2014b}, as well as the emergence of dynamic states referred to as 
{\em chimera states} in current literature \cite{Lazarides2015b,Hizanidis2016a}, 
have been demonstrated numerically in SQUID metamaterial models. Those 
counter-intuitive dynamic states have been discovered numerically in rings of 
identical phase oscillators\cite{Kuramoto2002} (see Ref. \cite{Panaggio2015} for 
a review).

\begin{figure}[!h]
\includegraphics[angle=0, width=0.85 \linewidth]{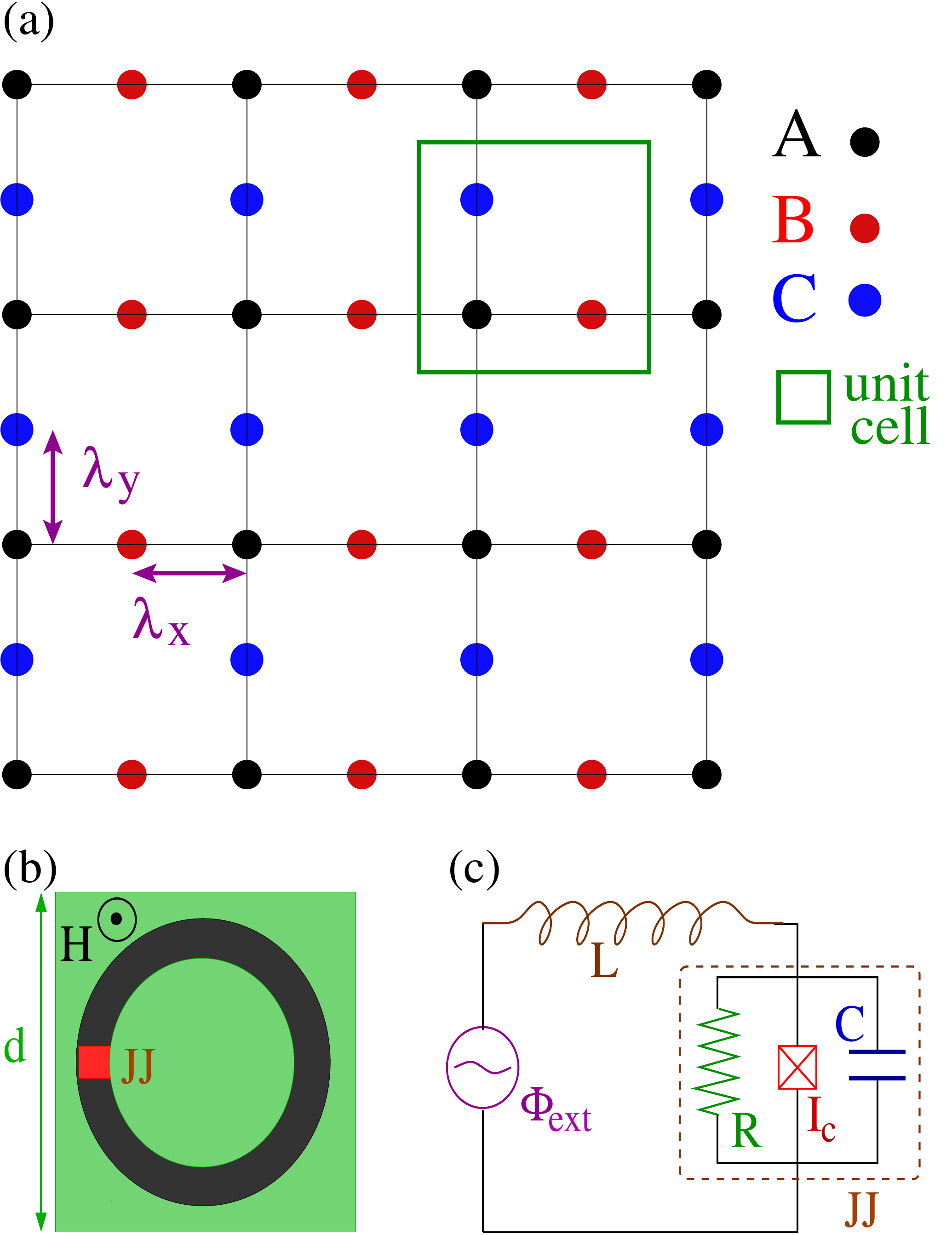}
\caption{(Color online)
(a) Schematic of a Lieb lattice; each unit cell (green square) has a corner 
    SQUID (black) and two edge SQUIDs (red and blue). The nearest-neighbor 
    coupling coefficients are indicated as $\lambda_x$ and $\lambda_y$; 
(b) Schematic of a single SQUID;
(c) Equivalent electrical circuit for a dissipative-driven SQUID.
\label{fig1}
}
\end{figure}

Experimental and theoretical investigations on SQUID metamaterials have been 
limited to quasi - one-dimensional (1D) lattices and two-dimensional (2D) 
tetragonal lattices. However, different arrangements of SQUIDs on the plane can 
be realized which may also give rise to novel band structures; for example, the 
arragnement of SQUIDs on a {\em line-centered tetragonal} (Lieb) lattice, which 
is described by three sites in a square unit cell (Fig. \ref{fig1}a), gives rise 
to a frequency spectrum featuring a Dirac cone intersected by a topological flat 
band. Such a {\em SQUID Lieb metamaterial} (SLiMM) supports compact flat-band 
localized states \cite{Lazarides2017}, very much alike to those observed in 
photonic Lieb lattices \cite{Vicencio2015,Mukherjee2015a}. Here, the existence 
of simultaneously stable excitations of the form of dissipative Discrete 
Breathers (DBs) is demonstrated numerically for a SLiMM which is driven by a 
sinusoidal flux field and it is subjected to dissipation. DBs are spatially 
localized and time-periodic excitations \cite{Flach2008a,Flach2012} whose 
existence has been proved rigorously for nonlinear Hamiltonian networks of 
weakly coupled oscillators \cite{Mackay1994,Aubry1997}. They actually have been 
observed in several physical systems such as Josephson ladders \cite{Binder2000} 
and Josephson arrays\cite{Trias2000}, micromechanical oscillator arrays 
\cite{Sato2003}, proteins \cite{Edler2004}, and antiferromagnets 
\cite{Schwarz1999}. From the large volume of research work on DBs, only a very 
small fraction is devoted to dissipative breathers, e.g., in Josephson ladders
\cite{Martinez1999,Martinez2003}, Frenkel-Kontorova lattices 
\cite{Marin2001,Martinez2003}, 2D Josephson arrays \cite{Mazo2002}, nonlinear 
metallic metamaterials \cite{Lazarides2006}, and 2D tetragonal SQUID 
metamaterials \cite{Lazarides2008a}. These excitations emerge through a delicate 
balance of input power and intrinsic losses. Dissipative beathers in Josephson 
arrays and ladders are reviewed in Ref. \cite{Mazo2003}; for a more general
review, see \cite{Flach2008b}. Note that dissipative breathers may exhibit 
richer dynamics than their Hamiltonian counterparts including quasiperiodic
\cite{Martinez2003} and chaotic \cite{Martinez1999,Martinez2003} behavior.
Moreover, simple 1D and 2D tetragonal lattices are considered in most works, 
except, e.g., those on moving DBs in a 2D hexagonal lattice \cite{Marin1998}, 
on DBs in cuprate-like lattices \cite{Marin2001b}, and on long-lived DBs in 
free-standing graphene (honeycomb lattice) \cite{Fraile2016}.

In the following, the dynamic equations for the fluxes through the loops of the 
SQUIDs of a SLiMM are quoted. Then, a typical snaking bifurcation curve of the 
flux amplitude as a function of the driving frequency for a single SQUID is 
presented, and its use for the construction of trivial dissipative DB 
configurations is explained. The existence of simultaneously stable dissipative 
DBs (from hereafter multistable DBs) at a frequency close to that of the 
single-SQUID resonance, is demonstrated. Bifurcation curves for the multistable 
DB amplitudes with varying the external flux field amplitude, the coupling 
coefficients, and the frequency of the driving flux field, are traced. For 
better understanding of those bifurcation diagrams, standard measures for energy 
localization and synchronization of coupled oscillators are calculated. 
Moreover, the existence of chimera states for appropriately chosen initial 
conditions is also demonstrated. Eventually, the wealth of dynamic behaviors 
that can be encountered in a SLiMM due to its lattice structure is indicated by 
the emergence of temporally chaotic states exhibiting a particular form of 
spatial coherence.

\section{Flux Dynamics Equations}
Consider the Lieb lattice of Fig. \ref{fig1}a, in which each site is occupied by 
a SQUID (Fig. \ref{fig1}b) modelled by the equivalent circuit shown in Fig. 
\ref{fig1}c; all the SQUIDs are identical, with each of them featuring a 
self-inductance $L$, a capacitance $C$, a resistance $R$, and a critical current 
of the Josephson junction $I_c$. The SQUIDs are magnetically coupled to their 
nearest-neighbors along the horizontal (vertical) direction through their mutual 
inductance $M_x$ ($M_y$). Assuming that the current in each SQUID is given by 
the resistively and capacitively shunted junction ($RCSJ$) model 
\cite{Likharev1986}, the dynamic equations for the fluxes through the loops of 
the SQUIDs are \cite{Lazarides2017}
\begin{eqnarray}
\label{1}
  L C \frac{d^2 \Phi_{n,m}^A}{dt^2} +\frac{L}{R} \frac{d \Phi_{n,m}^A}{dt}
   +L I_c \sin\left( 2\pi \frac{\Phi_{n,m}^A}{\Phi_0} \right) +\Phi_{n,m}^A 
\nonumber \\
  = \lambda_x \left( \Phi_{n,m}^B +\Phi_{n-1,m}^B \right)
   +\lambda_y \left( \Phi_{n,m}^C +\Phi_{n,m-1}^C \right) 
\nonumber \\
   +[1-2(\lambda_x +\lambda_y)] \Phi_{e} , \\
\label{2}
  L C \frac{d^2 \Phi_{n,m}^B}{dt^2} +\frac{L}{R} \frac{d \Phi_{n,m}^B}{dt}
   +L I_c \sin\left( 2\pi \frac{\Phi_{n,m}^B}{\Phi_0} \right) +\Phi_{n,m}^B 
\nonumber \\
  = \lambda_x \left( \Phi_{n,m}^A +\Phi_{n+1,m}^A \right) 
   +( 1-2 \lambda_x ) \Phi_{e} , \\ 
\label{3}
  L C \frac{d^2 \Phi_{n,m}^C}{dt^2} +\frac{L}{R} \frac{d \Phi_{n,m}^C}{dt}
   +L I_c \sin\left( 2\pi \frac{\Phi_{n,m}^C}{\Phi_0} \right) +\Phi_{n,m}^C 
\nonumber \\
  = \lambda_y \left( \Phi_{n,m}^A +\Phi_{n,m+1}^A \right) 
   +( 1-2 \lambda_y ) \Phi_{e} ,
\end{eqnarray}
where $\Phi_{n,m}^k$ is the flux through the loop of the SQUID of kind $k$ in 
the $(n,m)$th unit cell ($k=A$, $B$, $C$, the notation is as in Fig. 
\ref{fig1}a), $I_{n,m}^k$ is the current in the SQUID of kind $k$ in the 
$(n,m)$th unit cell, $\Phi_0$ is the flux quantum, $\lambda_x =M_x /L$ 
($\lambda_y =M_y /L$) is the coupling coefficient along the horizontal 
(vertical) direction, $t$ is the temporal variable, and 
$\Phi_{e} =\Phi_{ac} \, \cos( \omega t )$ is the external flux due to a 
sinusoidal magnetic field applied perpendicularly to the plane of the SLiMM. The 
subscript $n$ ($m$) runs from $1$ to $N_x$ ($1$ to $N_y$), so that 
$N=N_x \, N_y$ is the number of unit cells of the SLiMM (the number of SQUIDs is 
$3 N$).

Using the relations $\tau =\omega_{LC} t$, $\phi_{n,m}^k={\Phi_{n,m}^k} / {\Phi_0}$,
and $\phi_{ac}={\Phi_{ac}} / {\Phi_0}$, where $\omega_{LC} = {1}/{\sqrt{LC}}$ is the 
inductive-capacitive ($L C$) SQUID frequency, Eqs. (\ref{1})-(\ref{3}) can be 
normalized as
\begin{eqnarray}
\label{6}
{\cal L}\phi_{n,m}^A
  = \lambda_x \left( \phi_{n,m}^B +\phi_{n-1,m}^B \right) 
   +\lambda_y \left( \phi_{n,m}^C +\phi_{n,m-1}^C \right) 
\nonumber \\
   +[1-2(\lambda_x +\lambda_y)] \phi_e (\tau) , \\
\label{7}
{\cal L}\phi_{n,m}^B
  = \lambda_x \left( \phi_{n,m}^A +\phi_{n+1,m}^A \right) 
   +( 1-2 \lambda_x ) \phi_e (\tau) , \\ 
\label{8}
{\cal L}\phi_{n,m}^C
  = \lambda_y \left( \phi_{n,m}^A +\phi_{n,m+1}^A \right) 
   +( 1-2 \lambda_y ) \phi_e (\tau) ,
\end{eqnarray}
where
\begin{equation}
\label{8.2}
  \beta =\frac{L\, I_c}{\Phi_0} =\frac{\beta_L}{2\pi} ~~~ {\rm and} ~~~
  \gamma= \omega_{LC} \frac{L}{R}
\end{equation}
is the SQUID parameter and the dimensionless loss coefficient, respectively,
$\phi_e (\tau) = \phi_{ac} \cos(\Omega \tau)$ is the external flux of frequency 
$\Omega ={\omega} / {\omega_{LC}}$ and amplitude $\phi_{ac}$, and $\cal L$ is an 
operator such that
\begin{equation}
\label{9}
  {\cal L}\phi_{n,m}^k =\ddot{\phi}_{n,m}^k +\gamma \dot{\phi}_{n,m}^k +\phi_{n,m}^k
                        +\beta \sin\left( 2\pi \phi_{n,m}^k \right) . 
\end{equation}
The overdots on $\phi_{n,m}^k$ denote differentiation with respect to $\tau$.

The SQUID parameter and the loss coefficient used in the simulations have been 
chosen to be the same as those provided in the Supplemental Material of Ref. 
\cite{Zhang2015} for a $11 \times 11$ SQUID metamaterial, i.e., $\beta_L =0.86$
and $\gamma =0.01$. These values result from Eq. (\ref{8.2}) with $L=60~pH$, 
$C=0.42~pF$, $I_c =4.7~\mu A$, and subgap resistance $R=500$ Ohms. The value of
the coupling between neighboring SQUIDs has been chosen to be 
$\lambda_x =\lambda_y =-0.02$, as it has been estimated for a $27\times 27$
SQUID metamaterial in the experiments of Ref. \cite{Trepanier2013}.  
These experiments were performed with a specially designed set up which allows 
for the application of uniform ac driving and/or dc bias fluxes 
\cite{Trepanier2013,Zhang2015} as well as dc flux gradients \cite{Trepanier2017}
to the SQUID metamaterials which are placed into a waveguide.
In the simulations in the next Sections, the described effects can be identified
within the experimentally accessible range of $\phi_{ac}$ which spans the 
interval $0.001 - 0.1$ \cite{Zhang2015}. Furthermore, the SLiMM is chosen to 
have $16 \times 16$ unit cells, so that its size is comparable with that of the
$27 \times 27$ SQUID metamaterial investigated in Refs. 
\cite{Trepanier2013,Trepanier2017}.


\begin{figure}[!t]
\includegraphics[angle=0, width=0.85 \linewidth]{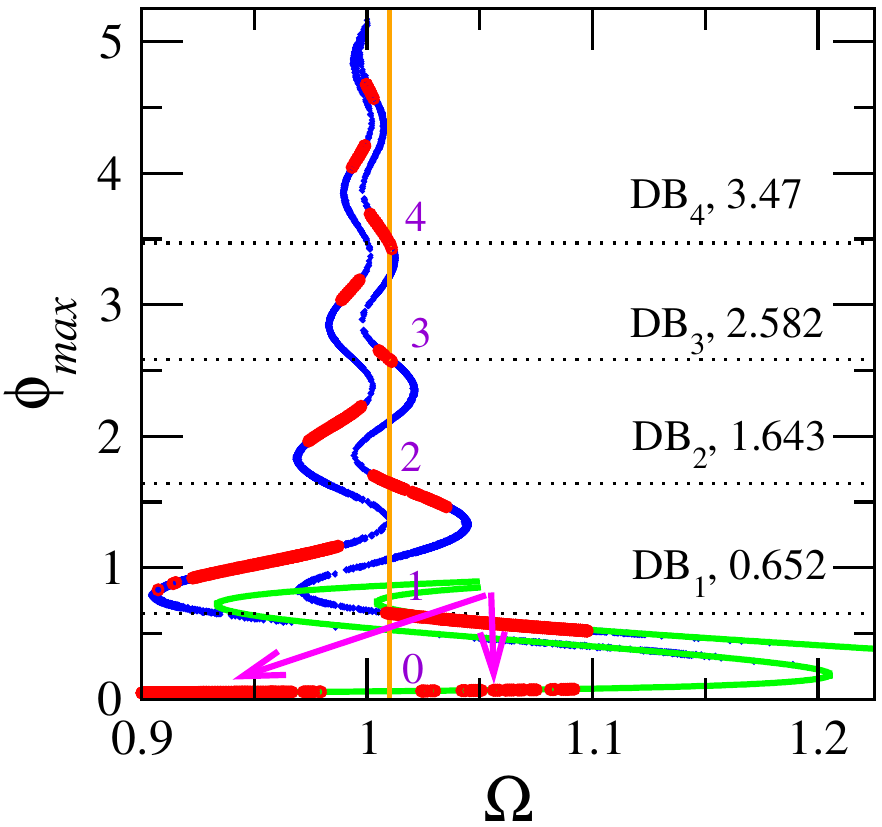} 
\caption{(Color online)
 The snaking flux amplitude $\phi_{max}$ - driving frequency $\Omega$ curve for 
 a single SQUID with $\beta_L =0.86$ and $\phi_{ac} =0.05$ (blue curves). The 
 green curves are calculated from Eq. (\ref{13}). The vertical orange line is at 
 frequency $\Omega=1.01$. The red symbols superposed on some branches of the 
 $\phi_{max}$ - $\Omega$ curve, are the amplitudes of {\em stable} dissipative 
 discrete breather families (except the ones indicated by the arrows, see text).
\label{fig2}
}
\end{figure}

\section{Single SQUID Resonance and Multistable Dissipative Breathers}
In a single SQUID driven with a relatively high amplitude field $\phi_{ac}$, 
strong nonlinearities shift the resonance frequency from $\Omega =\Omega_{SQ}$ 
to $\Omega \sim 1$, i.e., to the $LC$ frequency $\omega_{LC}$. Moreover, the 
curve for the oscillation amplitude of the flux through the loop of the SQUID 
$\phi_{max}$ as a function of the driving frequency $\Omega$ (SQUID resonance 
curve), aquires a snaking form as that shown in Fig. \ref{fig2} (blue) 
\cite{Hizanidis2016a}. That curve is calculated from the normalized single SQUID 
equation
\begin{equation}
\label{10}
  \ddot{\phi} +\gamma \dot{\phi} +\beta \sin\left( 2\pi \phi \right) +\phi 
    =\phi_{ac} \cos(\Omega \tau) ,
\end{equation}
for the flux $\phi$ through the loop of the SQUID. The curve "snakes" back and 
forth within a narrow frequency region via succesive saddle-node bifurcations 
(occuring at those points for which $d\Omega / d\phi_{max} =0$). The many 
branches of the resonance curve have been traced numerically using Newton's 
method; the stable branches are those which are partially covered by the red 
circles. An approximation to the resonance curve for $\phi_{max} \ll 1$ is given 
by \cite{Hizanidis2016a} 
\begin{eqnarray}
   \Omega^2 =\Omega_{SQ}^2 \pm \frac{\phi_{ac}}{\phi_{max}}  \nonumber \\
\label{13}
   -\beta_L \phi_{max}^2 
    \{ a_1 -\phi_{max}^2 [ a_2 -\phi_{max}^2 ( a_3 -a_4 \phi_{max}^2 )] \} ,
\end{eqnarray}
where $a_1 = \pi^2 /2$, $a_2 = \pi^4 /12$, $a_3 = \pi^6 /144$, and 
$a_4 = \pi^8 /2880$, which implicitly provides $\phi_{max} (\Omega)$. The 
approximate curves Eq. (\ref{13}) are shown in Fig. \ref{fig2} in green color; 
they show excellent agreement with the numerical snaking resonance curve for 
$\phi_{max} \lesssim 0.6$. The vertical orange segment at $\Omega =1.01$ 
intersects the resonance curve at several $\phi_{max}$ points; five of those, 
numbered on Fig. \ref{fig2} with consecutive integers from $0$ to $4$, 
correspond to stable solutions of the single SQUID equation. These five (5) 
solutions, which can be denoted as $(\phi_i, \dot{\phi_i})$ with 
$i=0, 1, 2, 3, 4$, are used for the construction of four (4) trivial dissipative 
DB configurations. Note that the flux amplitude $\phi_{max}$ of these five 
solutions increases with increasing $i$. For constructing a (single-site) 
trivial dissipative DB, two simultaneously stable solutions are first 
identified, say $(\phi_0, \dot{\phi_0})$ ($0$) and 
$(\phi_1, \dot{\phi_1})$ ($1$), with low and high flux amplitude $\phi_{max}$, 
respectively. Then, one of the SQUIDs at $(n, m)=(n_e =N_x/2, m_e =N_y/2)$ 
(hereafter referred to as the central DB site, which also determines the 
location of the DB) is set to the high amplitude solution $1$, while all the 
other SQUIDs of the SLiMM (the background) are set to the low amplitude solution 
$0$. In order to numerically obtain a dissipative DB, that trivial DB 
configuration is used as initial condition for the time-integration of Eqs. 
(\ref{6})-(\ref{8}); then, a stable dissipative DB (denoted as DB$_1$) is formed 
after integration for a few thousand time units. Three (3) more trivial 
dissipative DBs can be constructed similarly, e.g. by setting the central DB 
site to the solution $2$, $3$, or $4$, and the background to the solution $0$. 
Then, by integrating Eqs. (\ref{6})-(\ref{8}) using as initial conditions these 
trivial DB configurations, three more stable dissipative DBs are obtained 
numerically (denoted as DB$_2$, DB$_3$, and DB$_4$, respectively). These four 
dissipative DBs are simultaneously stable and oscillate with the driving 
frequency $\Omega =1.01$.

\begin{figure}[!t]
\includegraphics[angle=0, width=0.9 \linewidth]{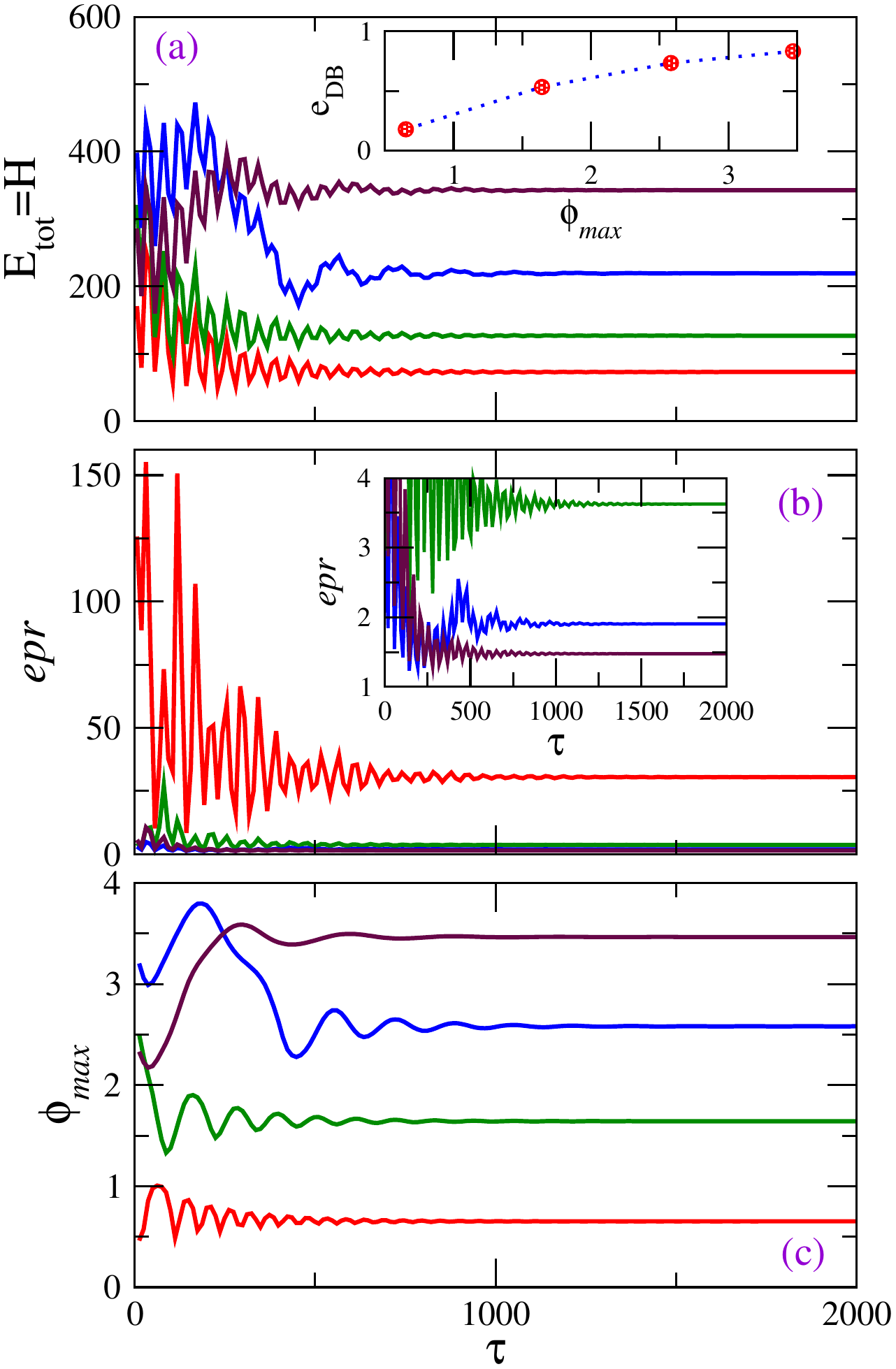} 
\caption{(Color online)
(a) The total energy $E_{tot} =H$ of the SQUID Lieb metamaterial as a function 
    of $\tau$ for $N_x =N_y =16$, $\beta_L =0.86$, $\lambda_x =\lambda_y =-0.02$, 
    $\gamma=0.01$, $\Omega =1.01$, $\phi_{ac} =0.05$, and four (4) initial 
    conditions - trivial breather configurations.
    Inset: The ratio $e_{DB} \equiv H_{n=n_e,m=m_e} / H$ as a function of the 
    steady-state dissipative breather amplitude $\phi_{max}$ for the four 
    multistable dissipative breathers. The blue-dotted curve is a guide to the 
    eye. 
(b) The energetic participation ratio $\it epr$ as a function of $\tau$ for the 
    four initial conditions - trivial breather configurations. 
    Inset: The $\it epr$ as a function of $\tau$ for the trivial breather 
    configurations leading to the three more localized dissipative breathers.
(c) The amplitude of the four multistable dissipative breathers $\phi_{max}$ as 
    a function of $\tau$. The asymptotic values of $\phi_{max}$ have been used 
    in the inset in (a).
\label{fig3}
}
\end{figure}

The Hamiltonian (total energy) for the SLiMM descibed by Eqs. 
(\ref{6})-(\ref{8}) for $\gamma=0$ is given by
\begin{equation}
\label{15}
  H =\sum_{n,m} H_{n,m} ,
\end{equation}
where the Hamiltonian (energy) density, $H_{n,m}$, is
\begin{eqnarray}
\label{16}
  H_{n,m} =\frac{\pi}{\beta} \sum_{k} \left[ 
           \left( q_{n,m}^k \right)^2 +\left( \phi_{n,m}^k -\phi_{e} \right)^2 \right]
\nonumber \\
  -\sum_{k} \cos\left( 2\pi \phi_{n,m}^k \right) 
\nonumber \\
  -\frac{\pi}{\beta} \{
   \lambda_x [ (\phi_{n,m}^A -\phi_{e})(\phi_{n-1,m}^B -\phi_{e}) 
\nonumber \\
             +2(\phi_{n,m}^A -\phi_{e})(\phi_{n,m}^B -\phi_{e})  
\nonumber \\
              +(\phi_{n,m}^B -\phi_{e})(\phi_{n+1,m}^A -\phi_{e}) ]
\nonumber \\
  +\lambda_y [ (\phi_{n,m}^A -\phi_{e})(\phi_{n,m-1}^C -\phi_{e})
\nonumber \\
             +2(\phi_{n,m}^A -\phi_{e})(\phi_{n,m}^C -\phi_{e})
\nonumber \\
             +(\phi_{n,m}^C -\phi_{e})(\phi_{n,m+1}^A -\phi_{e})] \} ,
\end{eqnarray}
where $q_{n,m}^k =\frac{d\phi_{n,m}^k}{d\tau}$ is the normalized instantaneous 
voltage across the Josephson junction of the SQUID in the $(n,m)$th unit cell of 
kind $k$. Both $H$ and $ H_{n,m}$ are normalized to the Josephson energy, $E_J$. 
Two more quantities are also defined; the {\em energetic participation ratio} 
\cite{deMoura2003,Laptyeva2012}
\begin{equation}
\label{17}
   epr =\left[ \sum_{n,m} \left( \frac{H_{n,m}}{H} \right)^2 \right]^{-1} ,
\end{equation} 
which is a measure of localization (it roughly measures the number of the most 
strongly excited unit cells), and the complex {\em synchronization parameter}
\begin{equation}
\label{18}
   \Psi =\frac{1}{3 N}  \sum_{n,m,k} e^{2\pi i \phi_{n,m}^k} ,
\end{equation} 
which is a spacially global measure of synchronization for coupled oscillators; 
its magnitude $r (\tau)=|\Psi (\tau)|$ ranges from zero (completely 
desynchronized solution) to unity (completely synchronized solution).

\begin{figure*}[!t]
\includegraphics[angle=0, width=0.96 \linewidth]{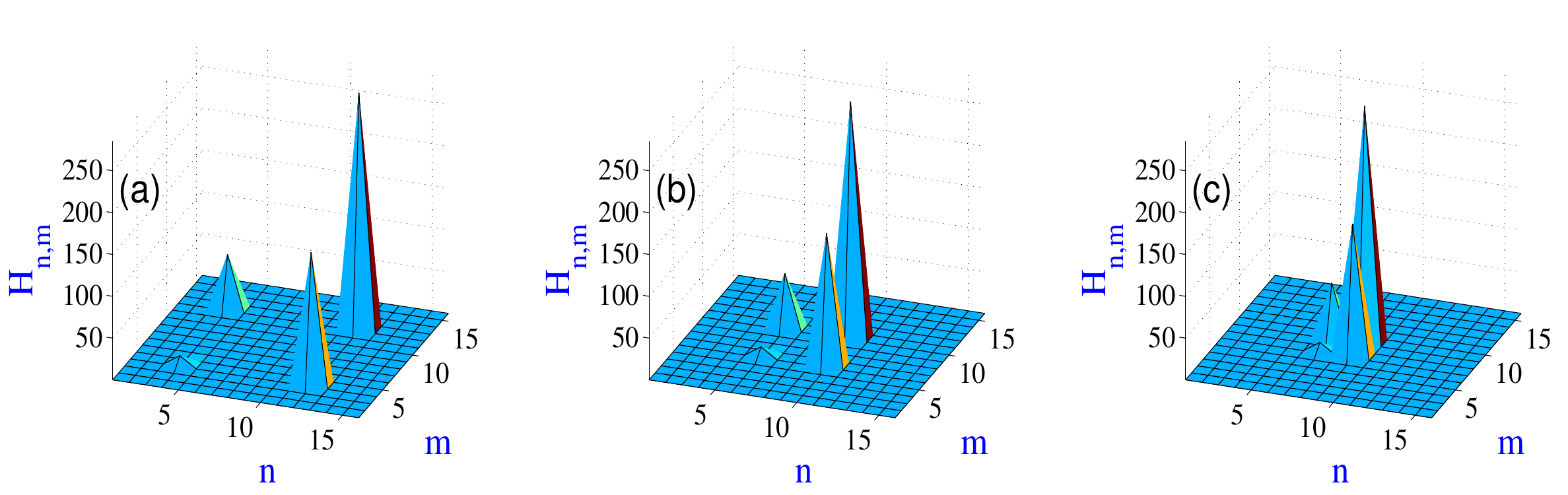} 
\caption{(Color online)
 The energy density $H_{n,m}$ of the SQUID Lieb metamaterial on the $n-m$ plane, 
 in which four dissipative breathers exist simultaneously, for $N_x =N_y =16$, 
 $\beta_L =0.86$, $\gamma=0.01$, $\lambda_x =\lambda_y =-0.02$, $\Omega =1.01$, 
 $\phi_{ac} =0.05$, and different separations. The central breather sites for 
 DB$_1$, DB$_2$, DB$_3$, and DB$_4$, are located on a square with vertices 
 respectively at 
 (a) $(n_e, m_e) =(4,4), ~(4,12), ~(12,4), ~(12,12)$;  
 (b) $(n_e, m_e) =(6,6), ~(6,10), ~(10,6), ~(10,10)$;  
 (c) $(n_e, m_e) =(7,7), ~(7,9), ~(9,7), ~(9,9)$.  
\label{fig4}
}
\end{figure*}

Eqs. (\ref{6})-(\ref{8}) implemented with periodic boundary conditions are 
initialized with the four trivial breather configurations and then integrated in 
time with a standard Runge-Kutta fourth order scheme. The temporal evolution of 
the total energy $H$ of the SLiMM, the energetic participation ratio $epr$, and 
the dissipative DB amplitude $\phi_{max}$ are shown for all cases in Fig. 
\ref{fig3}. After some oscillations during the initial stages of evolution, all 
curves flatten indicating that a steady state has been reached (after 
$\sim 1500$ time units of integration). As it can be observed, the SLiMM has 
higher energy for higher amplitude DBs $\phi_{max}$ (Fig. \ref{fig3}a). The 
steady-state values of $\phi_{max}$ for the four DBs can be seen in Fig. 
\ref{fig3}c; these values have been also used in the inset of Fig. \ref{fig3}a. 
In that inset, the ratio of the energy of the unit cell to which the central DB 
site belongs over the total energy of the SLiMM, i.e., $e_{DB} =H_{n_e,m_e} /H$, 
is shown for the four DBs. This ratio increases considerably with increasing DB 
amplitude. This is certainly compatible with Fig. \ref{fig3}b (see also the 
inset), in which $epr$ is plotted as a function of $\tau$, where apparently 
higher amplitude DBs provide more localized structures than lower amplitude ones. 

It is convenient to present the energy density $H_{n,m}$ profiles of the four 
dissipative DBs in one plot, as shown in Fig. \ref{fig4}. These profiles are 
obtained after $2000 ~T \simeq 12500$ time units of integration ($T=2\pi/\Omega$) 
using an appropriate initial condition which is a combination of the four trivial 
DB configurations. The difference between the three subfigures is in the 
distances between the central DB sites. Remarkably, the steady-state total 
energy of the SLiMM, $H=E_{tot}$, is the same in all the three cases and equal 
to $H=580.6$, indicating that the interaction between these DBs is almost 
negligible, even if they are located very closely (as in Fig. \ref{fig4}c).

\section{Bifurcations of Multistable Dissipative Breathers}
\begin{figure}[!t]
\includegraphics[angle=0, width=0.9 \linewidth]{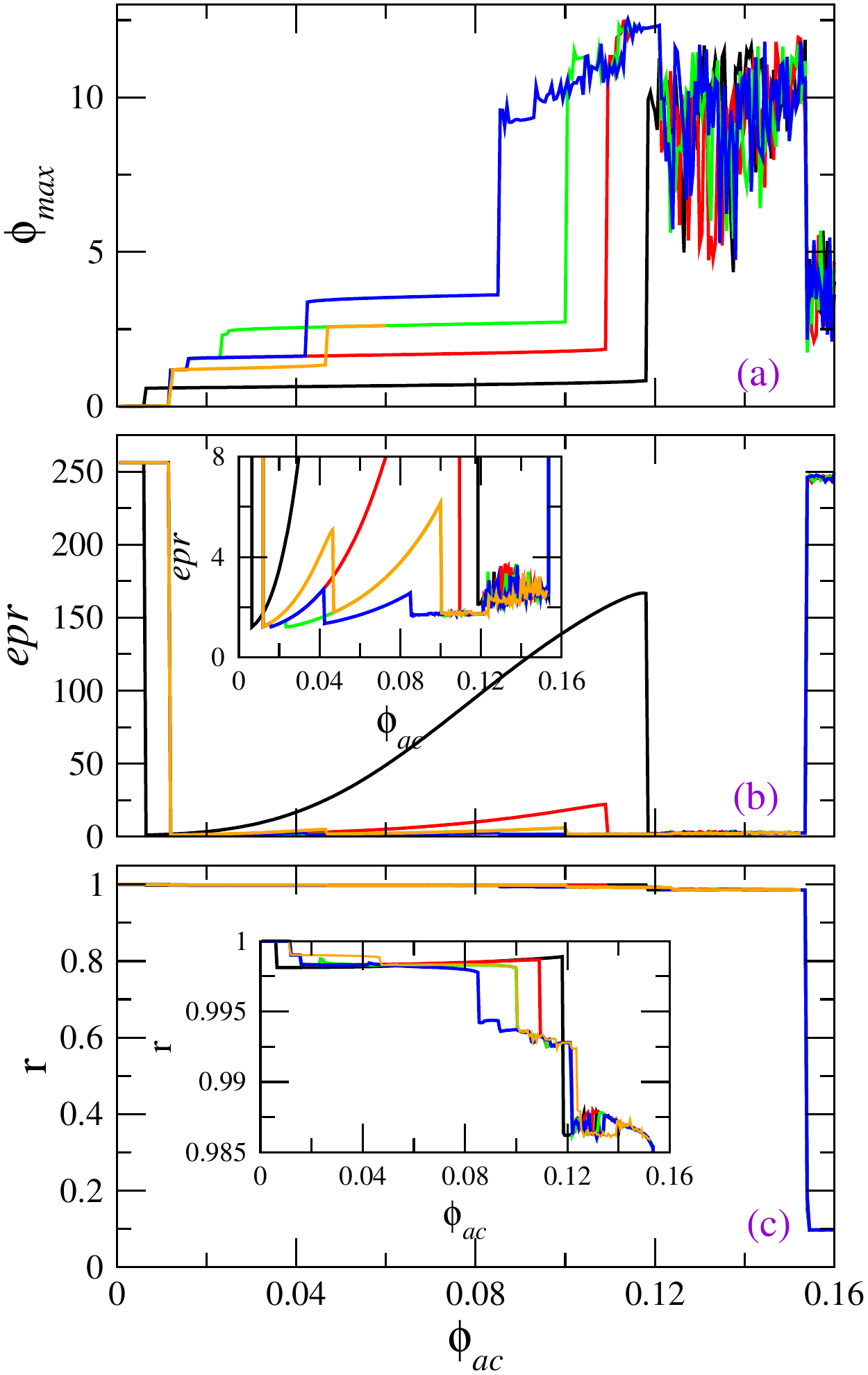} 
\caption{(Color online)
(a) The four dissipative breather amplitudes $\phi_{max}$ as a function of the driving 
    field amplitude $\phi_{ac}$, for $N_x =N_y =16$, $\beta_L =0.86$, $\gamma=0.01$, 
    $\Omega =1.01$, and $\lambda_x =\lambda_y =-0.02$. 
(b) The corresponding energetic participation ratios $epr$ as a function of 
    $\phi_{ac}$. Inset: Enlargement for low $epr$ values.
(c) The corresponding magnitudes of the synchronization parameter averaged over the 
    steady-state integration time $<r>_{int}$ as a function of $\phi_{ac}$. 
    Inset: Enlargement for values of $<r>_{int} \lesssim 1$.
\label{fig5}
}
\end{figure}

In this Section, the parameter intervals in which these four DBs are stable are 
determined; for this purpose, the steady-state DB amplitudes $\phi_{max}$ are 
calculated as a function of either the driving field amplitude $\phi_{ac}$, or the 
magnitude of the coupling coefficients for isotropic coupling $\lambda_x =\lambda_y$, 
or the driving frequency $\Omega$. First, $\phi_{max}$, the energetic participation
ratio $epr$, and the magnitude of the synchronization parameter averaged over the
steady-state integration time $\tau_{int} =2000 ~T$ time units (transients have been 
discarded), are calculated as a function of $\phi_{ac}$ (Fig. \ref{fig5}). In Fig. 
\ref{fig5}a, it can be seen that higher amplitude DBs remain stable for narrower 
intervals of $\phi_{ac}$. 
Interestingly, higher amplitude DBs may turn into lower amplitude ones even several 
times until they completely dissapear. As an example, we note that $DB_4$ (blue curve) 
which is stable approximately for $\phi_{ac}$ between $0.04$ and $0.085$, it 
transforms to a $DB_2$ for $\phi_{ac} <0.04$, and then to an even lower amplitude DB 
at $\phi_{ac} < 0.015$. The presence of the latter DB is rather unexpected, since it 
cannot be identified with one the four DB families under consideration. All the DBs 
disappear for $\phi_{ac} \lesssim 0.005$, since the nonlinearity is not strong enough 
to localize energy in the SLiMM. For $\phi_{ac}$ exceeding a critical value, which is 
higher for lower amplitude DBs (e.g., $0.085$ for $DB_4$ and $0.118$ for $DB_1$), all 
the four DBs turn into irregular multibreather states. 

In Figs. \ref{fig5}b and \ref{fig5}c the corresponding $epr$ and $<r>_{int}$ are 
presented as a function of $\phi_{ac}$. In Fig. \ref{fig5}b, it can be seen that when 
all the DBs dissapear for low $\phi_{ac}$, the SLiMM reaches a homogeneous  
state which is advocated by the large, close to the maximum possible value of 
$epr \simeq N =256$. In that case, $<r>_{int}$ is exactly unity (Fig. \ref{fig5}c) 
since the homogeneous state is synchronized. For high values of $\phi_{ac}$ 
($> 0.118$), where $\phi_{max}$ for all the four DBs varies irregularly with varying 
$\phi_{ac}$, the value of $epr$ can be used to distinguish between two different 
regimes: the first one from $\phi_{ac} \simeq 0.118$ to $0.154$, in which the low, 
fluctuating value of $epr$ suggests the existence of (possibly chaotic) multibreathers
(see also the inset of Fig. \ref{fig5}b), and the second from $\phi_{ac} \simeq 0.154$
to $0.16$, in which the high value of $epr$ ($\simeq 256$) suggests the existence of a
desynchronized state in which all the units cells are excited. For intermediate values
of $\phi_{ac}$, $epr$ generally increases with increasing $\phi_{ac}$; in particular, 
for DB$_1$ it increases to rather high values because of the relative enhancement of 
the oscillation amplitude of the background unit cells with respect to the central DB 
unit cell. However, this is not observed for the high amplitude DBs, for which the 
increase is either moderate (DB$_2$) or very small (DB$_3$, DB$_4$). It is also 
apparent that whenever a DB is tranformed to another, a small jump in $epr$ occurs 
(inset). Fig. \ref{fig5}c provides useful information on the synchronization of the 
various SLiMM states. For example, for $\phi_{ac} \simeq 0.154$ to $0.16$, $<r>_{int}$
falls off to very low values indicating desynchronization as mentioned above. 
For the values of $\phi_{ac}$ which provide stable single-site DBs that belong to one
of the four (4) families (as well as the fifth one which has appeared), the measure
$<r>_{int}$ is always very close to unity (inset); that occurs because all the 
"background" SQUIDs are oscillating in phase with the same amplitude, and only one 
SQUID (the central DB site) is oscillating with higher amplitude and opposite phase 
with respect to the others. For low $\phi_{ac}$, the SLiMM reaches a homogeneous 
state (the DBs have dissapeared) and then $<r>_{int}$ is exactly unity.  

\begin{figure}[!t]
\includegraphics[angle=0, width=0.9 \linewidth]{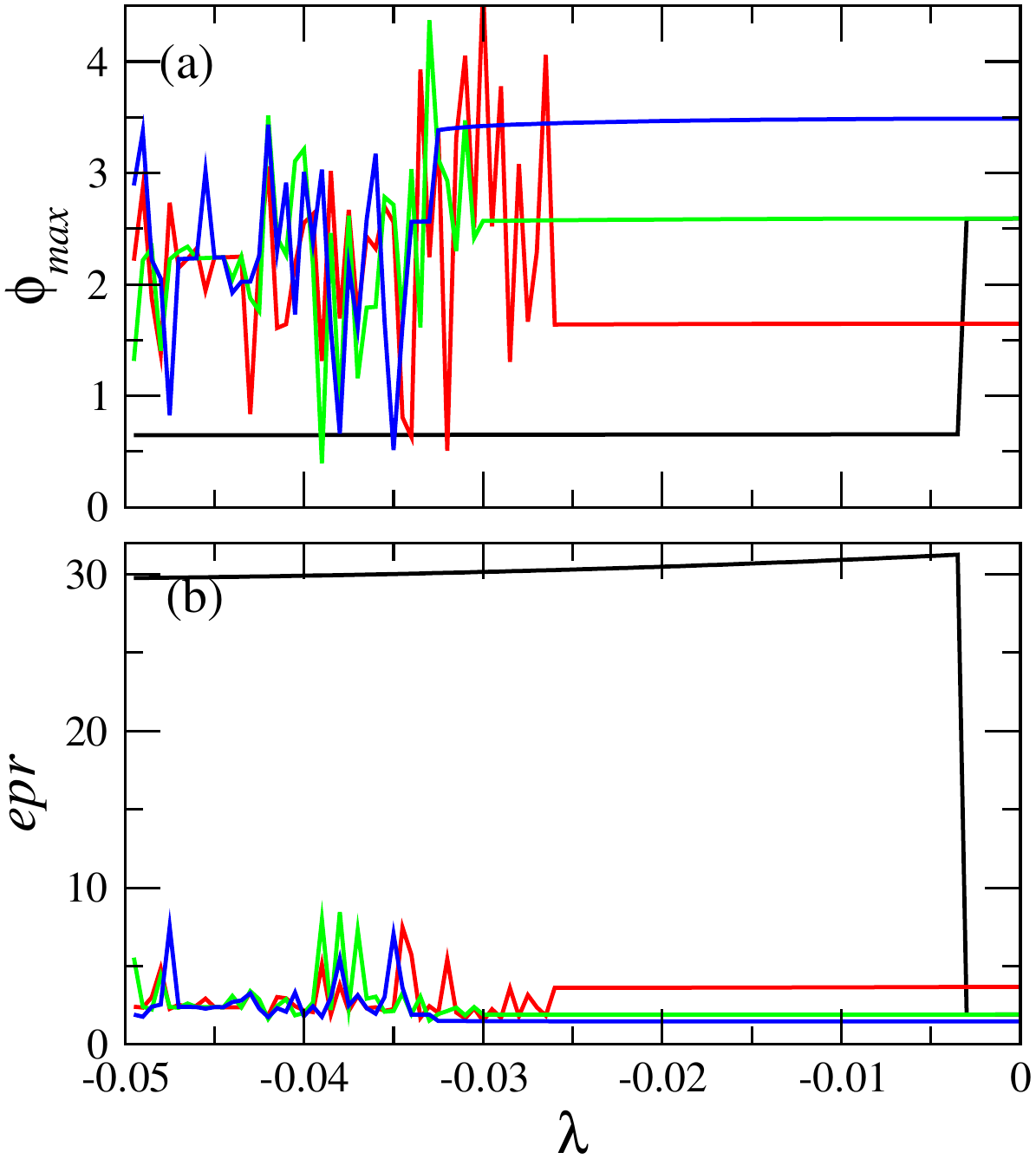} 
\caption{(Color online)
(a) The four dissipative breather amplitudes $\phi_{max}$ as a function of the 
    coupling coefficients in the case of isotropic coupling, 
    $\lambda =\lambda_x =\lambda_y$, for $N_x =N_y =16$, $\beta_L =0.86$, 
    $\gamma=0.01$, $\Omega =1.01$, and $\phi_{ac} =0.05$.  
(b) The corresponding energetic participation ratios $epr$ as a function of $\lambda$.
\label{fig6}
}
\end{figure}
The corresponding diagram of the DB flux amplitudes $\phi_{max}$ as a function of the 
coupling coefficients for isotropic coupling $\lambda =\lambda_x =\lambda_y$ is shown 
in Fig. \ref{fig6}. Remarkably, the four DBs maintain their amplitudes almost constant
for a substantial interval of $\lambda$ (Fig. \ref{fig6}a), i.e., from $\lambda =0$ to $-0.026$ which includes the estimated physically acceptable values for that system 
\cite{Trepanier2013,Trepanier2017}. The corresponding values of the $epr$ remain low, 
except for the lowest amplitude breather (DB$_1$), for which $epr \simeq 30$. 
Note that DB$_1$ dissapears for $\lambda > -0.003$ but it exists all the way down to 
$\lambda =-0.05$. For large magnitudes of $\lambda$, the amplitudes of the three high 
amplitude breathers (DB$_2$, DB$_3$, DB$_4$) vary irregularly with varying $\lambda$; 
however, as it can be observed in Fig. \ref{fig6}b, their $epr$ remains relatively 
low, indicating the spontaneous formation of multibreathers.

The bifurcation diagram of the DB amplitudes $\phi_{max}$ as a function of the driving 
frequency $\Omega$ is particularly interesting. This diagram has been superposed on 
the single SQUID resonance curve shown in Fig. \ref{fig2} as red circles. Notice that 
DB flux amplitudes (red circles) are very close to the corresponding flux amplitudes 
of single-SQUID stable solutions (which are covered by the red circles). All 
red-circled branches (except the lowest ones pointed by the arrows) correspond to 
stable DB families. The red-cirled branches indicated by the arrows correspond to 
almost homogeneous solutions which are not DBs. Note that different number of 
multistable dissipative DBs exists for different driving frequencies, depending on 
the broadness of the red-circled branches; for example, for $\Omega =1.01$
there are four simultaneously stable DBs, while for $\Omega =1.03$ there are two, and 
for $\Omega =1.07$ there is only one stable DB.   

\begin{figure}[!t]
\includegraphics[angle=0, width=0.9 \linewidth]{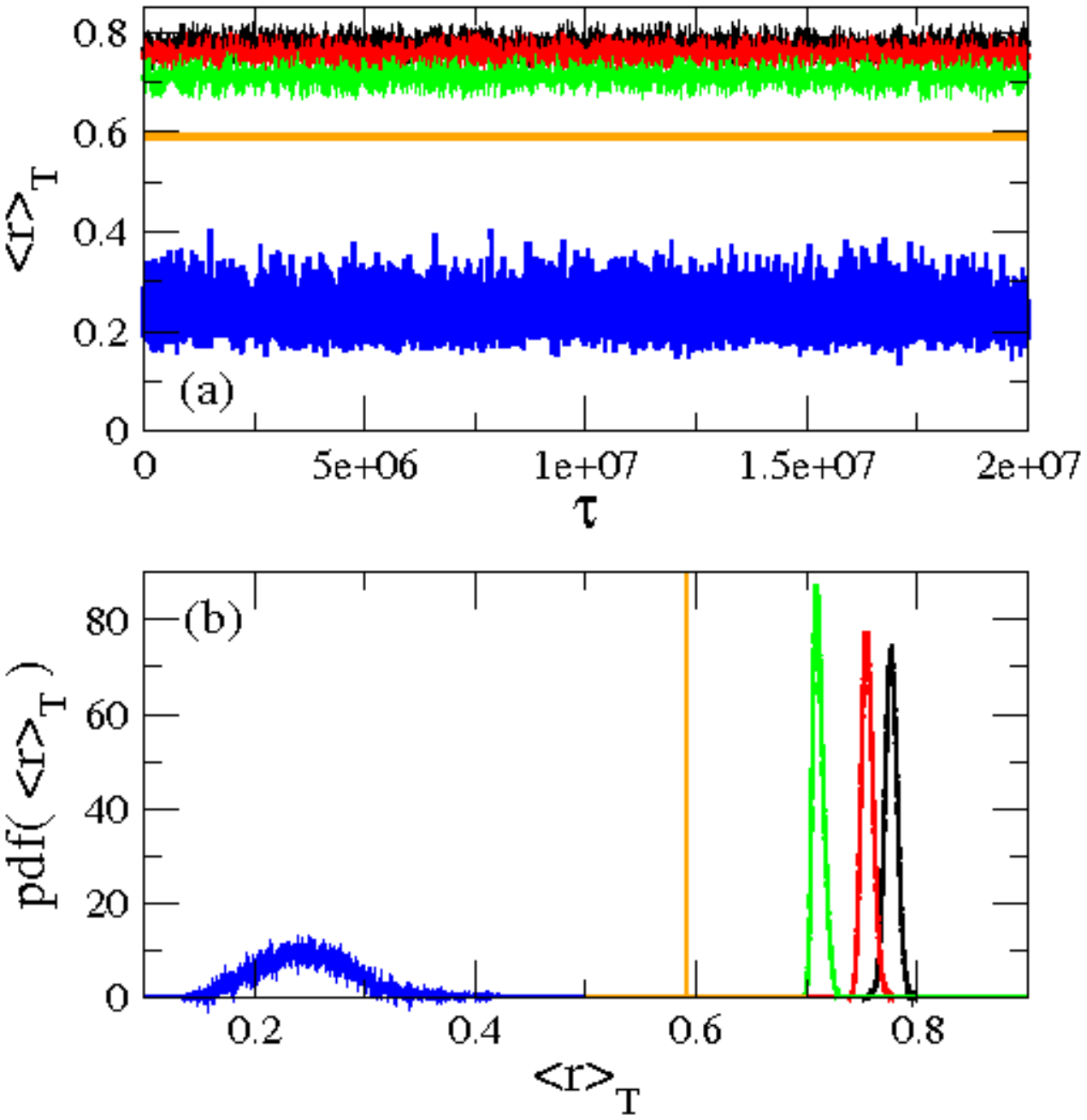} 
\caption{(Color online)
(a) The synchronization parameter averaged over the driving period $T$, $<r>_T$, as a 
    function of $\tau$ for $N_x =N_y =16$, $\beta=0.86$, $\gamma=0.01$, 
    $\lambda_x =\lambda_y =-0.02$, $\phi_{ac} =0.1$, and $\Omega =1.01$ (black); 
    $\Omega =1.02$ (red);$\Omega =1.03$ (green);$\Omega =1.04$ (blue); $\Omega =1.05$ 
    (orange).
(b) The corresponding probability distribution functions for the values of $<r>_T$,
    normalized to unity area. The actual peak of the orange curve, which is 
    practically a $\delta-$function is at $pdf( <r>_T ) =8000$.
\label{fig7}
}
\end{figure}

\section{Novel Dynamic SLiMM States} 
\begin{figure*}[!t]
\includegraphics[angle=0, width=0.96 \linewidth]{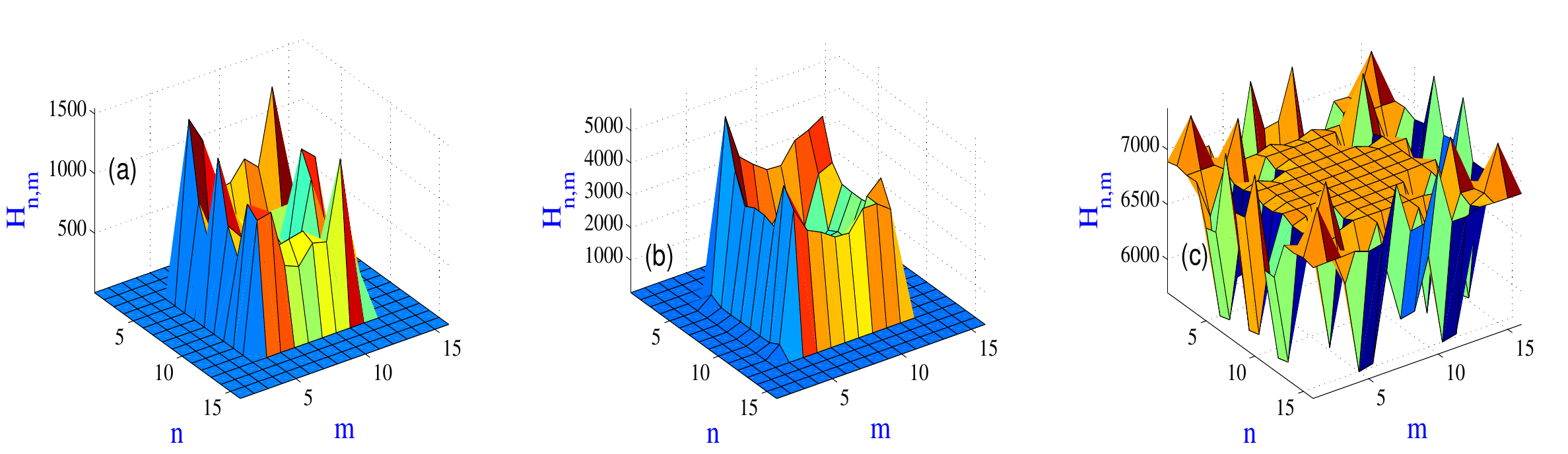} 
\caption{(Color online)
 The energy density $H_{n,m}$ of the SQUID Lieb metamaterial (energy per unit cell)
 on the $n-m$ plane, after integrating the dynamic equations for $10^7 ~T$ time units,
 for $N_x =N_y =16$, $\beta_L =0.86$, $\gamma=0.01$, $\lambda_x =\lambda_y =-0.02$, 
 $\phi_{ac} =0.1$, and (a) $\Omega =1.01$; (b) $\Omega =1.03$; (c) $\Omega =1.05$. 
 The value of the synchronization parameter averaged over the steady-state integration 
 time is $<r>_{int} \sim 0.77$, $\sim 0.71$, $\sim 0.59$, respectively. 
\label{fig8}
}
\end{figure*}

So far, we focused on the formation of single-site, dissipative DBs in a SLiMM, which 
can be generated through trivial DB configurations, and they are simultaneously 
stable. Beyond dissipative DB solutions, other interesting numerical solutions have 
been obtained; these solutions correspond to counter-intuitive dynamic states such as 
the so-called chimera states and a type of states that exhibit spatial homogeneity as 
well as chaotic evolution. Typical examples of such states, whose analysis requires 
futher work, are demonstrated here. First, a chimera state solution is illustrated 
which is generated from the following initial condition
\begin{eqnarray}
\label{19}
  \phi_{n,m}^k (\tau=0) =\left\{ \begin{array}{ll}
         0.5,  & \mbox{if~} N_x /4 +1 < n \leq 3 N_x/4    \\
               & \mbox{and~} N_y /4 +1 < m \leq 3 N_y/4;   \\
         0,    & \mbox{otherwise ,} \end{array} \right.  \\ 
\label{20}
 \dot{\phi}_{n, m}^k (\tau=0) =0 , \mbox{for any $n$, $m$} . 
\end{eqnarray}
With Eqs. (\ref{19}) and (\ref{20}) as initial conditions, Eqs. (\ref{6})-(\ref{8}) 
for the SLiMM are integrated in time. The magnitude of the synchronization parameter 
averaged over each driving period $T=2\pi/\Omega$, $<r>_T (\tau)$, is monitored in 
time and the results are shown in Fig. \ref{fig7}a for five different driving 
frequencies $\Omega$ close to unity. It can be seen that $<r>_T (\tau)$ is in all 
cases considerably less than unity, indicating significant desynchronization. The 
fluctuations, however, of $<r>_T (\tau)$ do not all have the same size. Specifically, 
for $\Omega=1.01$, $1.015$, and $1.02$ (black, red, and green curves, respectively), 
the fluctuations have roughly the same size. For $\Omega=1.025$ (blue curve), the size 
of fluctuations is significantly higher, while for $\Omega=1.03$ (orange curve) the 
fluctuations are practically zero. This can be seen more clearly in Fig. \ref{fig7}b, 
in which the distributions $pdf( <r>_T )$ of the values of $<r>_T (\tau)$ are shown; 
the full-width half-maximum (FWHM) of the $pdf( <r>_T )$s, quantifies the level of 
metastability of chimera states \cite{Shanahan2010,Lazarides2015a}. A partially 
desynchronized dynamic state (i.e., with $<r>_T<1$ but practically zero fluctuations), 
is not a chimera state but a clustered state, i.e., a non-homogeneous state in which 
different groups of SQUIDs oscillate with different amplitudes and phases with respect 
to the driving field; however, the SQUID oscillators that belong to the same group are 
synchronized. Thus, as can be inferred from Fig. \ref{fig7} as well as by the 
inspection of the flux profiles at the end of integration time (not shown), the curves 
for $\Omega=1.01$, $1.015$, $1.02$, and $1.025$ (black, red, green, and blue curves, 
respectively), are indeed due to chimera states. The energy density profiles at the 
end of the integration time for $\Omega=1.01$, $1.03$, and $1.05$, are shown in Fig. 
\ref{fig8}. The first two are typical for chimera states, while the last one is 
typical for a clustered state. Note that the SQUIDs within the square in which the 
fluxes were initialized to a non-zero value, oscillate with high amplitude and they 
are not synchronized. The rest of the SQUIDs, i.e., outside that square, oscillate in 
phase and with the same (low) amplitude. Thus, from the initial condition Eqs. 
(\ref{19}) and (\ref{20}), different chimera states are obtained for different driving 
frequencies. These states differ in their asymptotic value of $<r>_T$ as well as the 
FWHM of their $pdf( <r>_T )$s which determines their metastability level. In Fig. 
\ref{fig8}c, on the other hand, one may distinguish easily groups of SQUIDs with the 
same amplitude. The SQUIDs that belong to such a group are synchronized together 
while the groups are not synchronized to each other. In this clustered state, all the 
SQUIDs are oscillating with high amplitude (note the energy scales). 

\begin{figure*}[!t]
\includegraphics[angle=0, width=0.85 \linewidth]{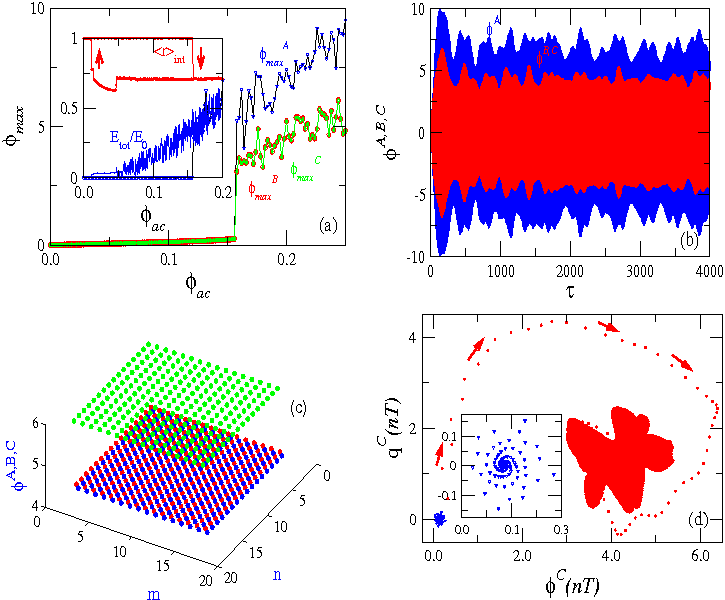} 
\caption{(Color online)
(a) The flux oscillation amplitudes $\phi_{max}^A$, $\phi_{max}^B$, and $\phi_{max}^C$
    of the SQUIDs of kind $A$ (blue), $B$ (red), and $C$ (green), respectively, of the 
    $(n_e,m_e)$th unit cell as a function of the driving field amplitude $\phi_{ac}$ 
    for $N_x =N_y =16$, $\beta_L =0.86$, $\gamma=0.01$, $\Omega =1.01$, and 
    $\lambda_x =\lambda_y =-0.02$.
    Inset: 
    The magnitude of the synchronization parameter averaged over the steady-state
    integration time $<r>_{int}$ (red) and the total energy of the SQUID Lieb 
    metamaterial divided by $E_0$, $E_{tot} / E_0$ (blue), as a function of 
    $\phi_{ac}$. Note the large hysteresis region in those curves.
(b) Time-dependence of $\phi^A$ (blue), $\phi^B$ (red), and $\phi^C$ (green), for 
    $\phi_{ac}=0.2$ and the other parameters as in (a).
(c) The fluxes $\phi^A$ (green), $\phi^B$ (red), and $\phi^C$ (blue), on the $n-m$ 
    plane for $\phi_{ac}=0.2$ and the other parameters as in (a).
(d) Stroboscopic plots of $\phi^C (nT) - q^C (nT)$, with $q^C \equiv \dot{\phi}^C$,
    for $\phi_{ac}=0.1$ (blue) and $\phi_{ac}=0.2$ (red). An enlargement of the
    period-1 attractor is shown in the inset. The red arrows are along the transient
    leading to the chaotic attractor. 
\label{fig9}
}
\end{figure*}
A family of novel solutions emerges through an order-to-chaos phase transition that 
is demonstrated for varying $\phi_{ac}$. Eqs. (\ref{6})-(\ref{8}) with periodic 
boundary conditions are integrated in time for $\phi_{ac}$ increasing from zero to 
higher values; the initial condition is homogeneous, i.e., 
$\phi_{n,m}^k (\tau=0) =\dot{\phi}_{n,m}^k (\tau=0) =0$ for any $n$, $m$, and $k$. The 
flux field amplitude $\phi_{ac}$ increases in small steps, and for each step the 
solution for the previous step is taken as the initial condition. For relatively low 
$\phi_{ac}$, the amplitudes of the oscillating fluxes through the loops of the SQUIDs 
have low values and they are very close to each other, i.e., 
$\phi_{max}^A \simeq \phi_{max}^B =\phi_{max}^C$ as shown in Fig. \ref{fig9}a
(all the SQUIDs of kind $k$ are oscillating with amplitude $\phi_{max}^k$, $k=A,B,C$). 
Actually, the difference between $\phi_{max}^A$ and $\phi_{max}^{B,C}$ is less than 
$1\%$ in this regime. Moreover, the fluxes in all kinds of SQUIDs are oscillating 
periodically in phase, and thus the degree of synchronization $<r>_{int}$ of these 
states is almost unity (see the upper branch of the red curve in the inset of Fig. 
\ref{fig9}a). That state is referred to as an {\em almost} homogeneous state in space.
In the inset of Fig. \ref{fig9}a, the total energy of the SLiMM $E_{tot}$ divided by 
$E_0=10^6$ is plotted as a function of $\phi_{ac}$; that energy increases smoothly 
with increasing $\phi_{ac}$ (lower branch of the blue curve in the inset of Fig. 
\ref{fig9}a). At a critical value of $\phi_{ac}$, $\phi_{ac}^c \simeq 0.155$, the 
situation changes drastically, as an abrupt increase of all the amplitudes 
$\phi_{max}^k$ occurs while their values become considerably different 
($\phi_{max}^A$ attains considerably larger values than $\phi_{max}^B =\phi_{max}^C$). Moreover, for $\phi_{ac} > \phi_{ac}^c$, the values of $\phi_{max}^k$s vary 
irregularly with increasing $\phi_{ac}$, although their average values as well as the 
difference between $\phi_{max}^A$ and $\phi_{max}^{B,C}$ increase (Fig. \ref{fig9}a). 
Also, at the phase transition point $\phi_{ac}^c$, the parameter $<r>_{int}$ abruptly 
jumps to a value which indicates significant desynchronization, $<r>_{int} \sim 0.7$; 
that value remains almost unchanged with further increasing $\phi_{ac}$ (inset). The 
variation of the total energy of the SLiMM $E_{tot}$ is similar to that of the 
variation of the $\phi_{max}^k$, i.e., it jumps abruptly to higher value at 
$\phi_{ac} =\phi_{ac}^c$ (inset). Recall that the above remarks hold for $\phi_{ac}$ 
increasing from zero to higher values. The corresponding curves for $<r>_{int}$ and 
$E_{tot}$ for $\phi_{ac}$ decreasing from $0.3$ to zero are also shown in the inset of Fig. \ref{fig9}a (lower branch of the red curve and upper branch of the blue curve, 
respectively). The "explosive" (first-order) character of the transition is clearly 
manifested by the presence of a large hysteresis region.

Consider again the case in which $\phi_{ac}$ increases from zero to higher values. 
In that case, the steady-states of the SLiMM are almost synchronized (almost 
spatially homogeneous) and temporally periodic for $\phi_{ac} < \phi_{ac}^c$. Note 
however that those states are exactly homogeneous at the unit cell level, i.e., that 
$\bar{\phi}_{n,m} =\sum_k {\phi}_{n,m}^k =c$ for any $n$ and $m$, with $c$ being a 
constant. For $\phi_{ac} > \phi_{ac}^c$ the SLiMM states acquire chaotic 
time-dependence, while they retain partial homogeneity and thus synchronization; that 
is, all the SQUIDs of kind $k$ are synchronized although they execute chaotic 
oscillations. Remarkably, at the unit cell level, even this state is spatially 
homogeneous. In Fig. \ref{fig9}b the time-dependence of the fluxes $\phi^A$, $\phi^B$,
and $\phi^C$ (identical for all the SQUID of kind $A$, $B$, and $C$, respectively, of 
the SLiMM) are plotted for $\phi_{ac} =0.2$ during a few thousands time-units. 
Apparently, the flux oscillations are irregular, indicating chaotic behavior which has
been checked to persists for very long times (note that $\phi^B =\phi^C$ due to the 
isotropic coupling). A flux profile for that state is shown in Fig. \ref{fig9}c, in 
which the spatial homogeneity within each sublattice of the SLiMM is apparent. Thus, 
large scale synchronization between oscillators in a chaotic state occurs in this 
case. In Fig. \ref{fig9}d, two stroboscopic plots in the reduced 
$\phi^C$ - $\dot{\phi}^C$ ($\dot{\phi}^C =q^C$) phase space are shown together for the
$C$ SQUID at the $(n_e, m_e)$th unit cell. In the first one (blue down-triangles, 
inset), the SLiMM is in an almost synchronized temporally periodic state 
($\phi_{ac} =0.1$); in the the second one (red circles), the SLiMM is in a partially 
synchronized (synchronization of the SQUIDs within each sublattice) temporally 
chaotic state ($\phi_{ac} =0.2$). Apparently, the trajectory in the reduced 
phase-space tends to a point in the former case, while it tends to a large area 
attractor in the latter. In Fig. \ref{fig9}d, the transients leading the trajectories 
to the one or the other attractor are also shown.

\section{Conclusions}
The existence of several regions in parameter space in which simultaneously stable
dissipative DBs in a dissipative SLiMM which is driven by a sinusoidal flux field has 
been demonstrated numerically. For that purpose, the dynamic equations Eqs. 
(\ref{6})-(\ref{8}) for the fluxes threading the loops of the SQUIDs are integrated in 
time with periodic boundary conditions. The initial conditions have been properly 
designed to provide trivial DB configurations using combinations of simultaneously 
stable solutions of the single-SQUID oscillator. For substantial nonlinearity excited 
from a flux field with relatively high amplitude, the single-SQUID resonance curve 
exhibits several simultaneously stable solutions at frequencies around resonance. That 
allows for the construction of several trivial DB configurations at some particular 
frequency; the subsequent temporal evolution through Eqs. (\ref{6})-(\ref{8}) results 
in multistable (co-existing) dissipative DBs. The bifucation diagrams for the 
calculated DB amplitudes as a function of $\phi_{ac}$, $\lambda$, and $\Omega$ have 
been presented, which reveal that multistability persists within substantial parameter 
intervals. For a better interpretation of those bifurcation diagrams, well-established 
measures for energy localization and synchronization of oscillators in discrete 
lattices were defined, and they were calculated for each dissipative DB. Remarkably, 
the interactions between co-existing DBs are very weak; no appreciable change in the 
total energy of the SLiMM has been observed even when the co-existing DBs are very 
close together. The bifurcation diagram of the dissipative DB amplitudes as a function 
of $\Omega$, shown as the branches formed by the red circles in Fig. \ref{fig2}, 
resembles the snaking bifurcation curves for spatially localized states in the 
Swift-Hohenberg equation \cite{Knobloch2008,Bergeon2008}; however, snaking bifurcation 
curves also occur in discrete problems \cite{Taylor2010,Prilepsky2012,Dean2015}. 
Interestingly, snaking bifurcation diagrams for chimera states have been obtained in 
the 1D extended Bogdanov-Takens lattice \cite{Clerc2016}.  

Besides single-site multistable dissipative DBs, two other types of dynamic states 
were demonstrated; chimera states, which can be generated in a SLiMM by appropriate 
choice of initial conditions, and spatially homogeneous (at the unit cell level) - 
temporally chaotic states. The existence of the former have been demonstrated in 1D 
SQUID metamaterials, and the mechanism for their generation through the {\em attractor
crowding} effect in coupled nonlinear oscillator arrays 
\cite{Wiesenfeld1989,Tsang1990} has been described
\cite{Lazarides2015b,Hizanidis2016a}. Similar chimera states are also expected to 
appear in SQUID metamaterials on 2D tetragonal lattices. The spatially homogeneous -
temporally chaotic states, however, are peculiar to the lattice geometry of the SLiMM,
and indicate the wealth of dynamic behaviors that may be encountered in that system. 
As $\phi_{ac}$ increases from zero, the SLiMM passes through states which are 
spatially (almost) homogeneous and temporally periodic. At a critical value of 
$\phi_{ac} =\phi_{ac}^c$ a transition occurs, and for $\phi_{ac} > \phi_{ac}^c$ the 
SLiMM passes through states in which all the SQUIDs of kind $k$ (i.e., the SQUIDs 
within each sublattice) have the same amplitude and they are synchronized together, 
while their time-dependence is chaotic! These states exhibit large-scale chaotic 
synchronization \cite{Pecora1990,Heagy1994}; notably, states with spatial coherence 
and temporal chaos have been obtained in coupled map lattices with asymmetric 
short-range couplings \cite{Aranson1992}. The order-to-chaos transition with 
hysteresis obtained here is similar to that demonstrated numerically and observed in 
laser-cooled trapped ions \cite{Hoffnagle1988}. When seen as a 
synchronization-desynchronization transition with hysteresis, it resembles the 
explosive first order transition to synchrony observed in electronic circuits 
\cite{Levya2012}.     

\section*{ACKNOWLEDGMENT}
This work is partially supported by the Ministry of Education and Science of the 
Russian Federation in the framework of the Increase Competitiveness Program of NUST 
"MISiS" (No. K2-2017-006), and by the European Union under project NHQWAVE 
(MSCA-RISE 691209).
NL gratefully acknowledges the Laboratory for Superconducting Metamaterials, NUST 
"MISiS" for its warm hospitality during visits.




\end{document}